\documentclass[]{pasj01}

\Received{}
\Accepted{}

\usepackage[switch,mathlines]{lineno}
\usepackage{color}

\usepackage{ulem}

\begin{document} 

\title{Polarized X--rays Correlated with Short--Timescale Variability of Cygnus X--1

\LETTERLABEL 
}
\author{Kaito \textsc{Ninoyu}\altaffilmark{1,}\thefootnote{*}}
\email{6223525@ed.tus.ac.jp}

\author{Yuusuke \textsc{Uchida}\altaffilmark{1,}\thefootnote{*}}
\email{yuuchida@rs.tus.ac.jp}

\author{Shinya \textsc{Yamada},\altaffilmark{2,}\thefootnote{*}}
\email{syamada@rikkyo.ac.jp}

\author{Takayoshi \textsc{Kohmura}\altaffilmark{1}}

\author{Taichi \textsc{Igarashi}\altaffilmark{2,3}}

\author{Ryota \textsc{Hayakawa}\altaffilmark{2,4}}

\author{Tenyo \textsc{Kawamura}\altaffilmark{2}}

\altaffiltext{1}{Department of Physics and Astronomy, Tokyo University of Science, 2641 Yamazaki, Noda, Chiba 278-8510, Japan}

\altaffiltext{2}{Department of Physics, Rikkyo University, 3-34-1 Nishi Ikebukuro, Toshima-ku, Tokyo 171-8501, Japan}

\altaffiltext{3}{Division of Science, National Astronomical Observatory of Japan, Tokyo, Japan}

\altaffiltext{4}{International Center for Quantum-field Measurement Systems for Studies of the Universe and Particles (QUP), KEK, 1-1 Oho, Tsukuba, Ibaraki 305-0801, Japan}


\maketitle

\begin{abstract}
We systematically investigate the variability of polarized X--rays on a timescale of a few seconds in the low/hard state of the black hole binary Cygnus X--1. The correlation between polarization degrees and angles with X--Ray intensity was analyzed using data collected by the Imaging X--ray Polarimetry Explorer (IXPE) in June 2022. Given that X--Ray variability in the low/hard state of Cygnus X--1 is non-periodic, flux peaks were aggregated to suppress statistical fluctuations. We divided the temporal profiles of these aggregated flux peaks into seven time segments and evaluated the polarization for each segment. The results reveal that the polarization degree was 4.6\%$\pm$1.2 and 5.3\%$\pm$1.2 before and after the peak, respectively, but decreased to 3.4\%$\pm$1.1 and 2.7\%$\pm$1.1 in the segments including and immediately following the peak. Furthermore, the polarization angle exhibited a slight shift from approximately 30$^{\circ}$ to $\sim$40$^{\circ}$ before and after the peak. These findings suggest that the accretion disk contracts with increasing X--Ray luminosity, and the closer proximity of the X--Ray emitting gas to the black hole may lead to reduced polarization.
\end{abstract}


\clearpage

\section{Introduction}
Cygnus X--1 (Cyg X--1) stands as one of the most celebrated black hole binary (BHB) systems. This system is composed of a black hole with a mass of $21.2$$\pm$2.2 solar masses and a companion blue giant star with a mass of $40.6^{+7.7}_{-7.1}$ solar masses, located at a distance of 2.2 kpc \citep{Miller-Jones2021}. The black hole accumulates material from its companion star, forming accretion flows that heat up to several million Kelvin. Cyg X--1 is known to display two distinctive spectral features \citep{Done2007}: the high/soft state and the low/hard state. In the high/soft state, radiation in the soft X--ray band is predominantly due to multi black-body radiation from the accretion disk (e.g., \cite{Tomsick2014, Walton2016}). Conversely, in the low/hard state, radiation is dominated by inverse Compton scattering in the corona (e.g., \cite{Makishima2008, Yamada2013Evidence}), characterized by a high-temperature electron cloud with $kT_\mathrm{e} \sim 100$~keV and their reflection components by the accretion disk including the fluorescence iron--K lines \citep{Fabian2012}. Spectral and timing analysis was used to investigate the structure and radiation mechanism of the accretion flows, in particular spectroscopy of the energy spectrum (e.g., \cite{Basak2017, KrawczynskiandBeheshtipour2022}), the time lag of the X--ray variability between different energy bands (e.g, \cite{Nowak1999a,Pottschmidt2000}) or the power spectra showing X--ray variability in frequency space (e.g., \cite{AxelssonandDone2018, MahmoudandDone2018}). Determining the detailed geometries has been difficult due to modeling degeneracies, requiring alternative approaches. 

The black hole accretion flow often emits X--rays with distinct polarization characteristics. Such X--ray polarization is believed to arise from multiple sources, including radiation from within the accretion disk \citep{Schnittman2009}, Compton scattering in the corona, radiation reflections oﬀ the accretion disk, and outflows \citep{Poutanen1996, Schnittman2010}. These polarized X--rays offer clues about the distribution of accretion flow close to the black hole and help elucidate the geometry of both the accretion disk and corona. The Eighth Orbiting Solar Observatory (OSO--8) detected the polarization of Cyg X--1 for the first time \citep{Long1980, Weisskopf1977}. After around forty years, PoGO+ observations revealed that the upper limit of polarization degree is 8.6\% and the polarization angle is parallel to the jet axis in the high energy band of 19--181~keV \citep{Chauvin2018}. A remarkable observation of Cyg X--1 using the Imaging X--ray Polarimetry Explorer (IXPE; \cite{Weisskopf2022}) in May 2022 revealed a polarization degree of $4.0\pm0.2$\% and a polarization angle of $-20^\circ.7\pm1^\circ.4$ (\cite{Krawczynski2022}, hereafter K22), which is higher than expectations of a polarization degree of $\sim1$\% \citep{KrawczynskiandBeheshtipour2022}. Notably, this is in alignment with the jet position angle \citep{Stirling2001} observed in the 2--8~keV range. Such findings suggest a corona distributed perpendicular to the direction of the jet and in alignment with the accretion disk. Interestingly, even this geometry of accretion disk and corona struggles to account for the high polarization of 4\% at an low inclination angle $27^\circ.1\pm0^\circ.8$ \citep{Orosz2011}, hinting at the need for a inclination angle exceeding $45^\circ$. As X--ray polarization offers a window into the dynamics around black holes, any rapid changes in the accretion flow's structure could lead to accompanying shifts in polarization patterns.

Cyg X--1 has consistently exhibited intensity variability on short duration spanning a few seconds particularly in its the low/hard state, as documented by prior studies such as \citet{Miyamoto1992}, Negoro et al. (\yearcite{Negoro1994}, hereafter N94), and Yamada et al. (\yearcite{Yamada2013Rapid}, hereafter Y13). Given that each observed peak in the X--ray variability tends to be of relatively low amplitude and non--period, and thus lacks sufficient statistical robustness, there is an imperative to aggregate these peaks for improved clarity. N94 introduced the ``shot analysis'' and applied it to the observation of Cyg X--1 with Ginga, and Y13 applied it to the observation of Cyg X--1 with Suzaku. They were found to be time-symmetric in terms of the characteristics of the intensity change from brightening to darkening. The hardness ratio of the N94 and Y13 profiles up to 60~keV and 200~keV, respectively, shows that the profiles become softer toward the peak and harder rapidly after the peak. Y13 revealed the variability of the Compton component parameters (the electronic temperature, optical depth and $y$--parameter) with an intensity variation and suggested that they can be explained by physical phenomena in the accretion flow at $\sim$1~s or less. 

To investigate the short--timescale polarization variability, we focused on analyzing the rapid intensity variability. In this letter, we show that we have explored the properties of the polarization variability by polarization detection method with aggregation of X--ray intensity peaks.

\section{Observation and Data Reduction}
\begin{figure*}[tb]
	\begin{center}
	  \includegraphics[width=\linewidth]{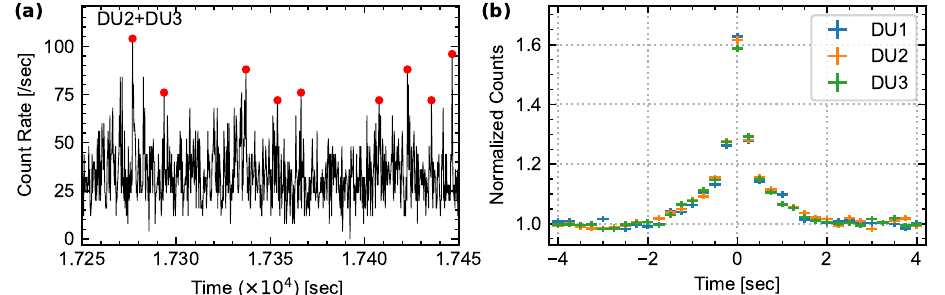} 
	\end{center}
	\caption{Combined light curve and profile of aggregated X--ray peaks across detector units. (a), Light curve constructed from the combined data of DU2 and DU3. The x-axis indicates the elapsed time since the beginning of the observation, and the y-axis shows the count rate with a temporal resolution of 0.25~s. Detected peaks of X--ray flux are emphasized in red. (b), Separate profiles for each detector unit: DU1 in blue, DU2 in orange, and DU3 in green.}
	\label{lightcurve_and_shotprofile}
\end{figure*}

IXPE can acquire enough statistics to detect the rapid variability of the polarization, thanks to its high sensitivity of polarization detection. Carrying three detector units (DUs), each paired with a Mirror Module Assembly, the design of IXPE is optimal for the X--ray polarimeters. The Gas Pixel Detectors (GPDs) on board each DUs achieve polarization for incident polarized X--ray photons in the energy band of 2--8~keV by measuring the electric field direction of each incident polarized X--ray photon, as detailed in \citet{Baldini2021}. For a comprehensive understanding of the polarization measurement principle employed by GPD, readers are directed to \citet{DiMarco2022}, \citet{Muleri2022} and \citet{Baldini2022}.

Cyg X--1 underwent observations via IXPE on six separate occasions between May 2022 and June 2024. For the purpose of this study, we focused on the low/hard state observations: specifically, the 246~ks observation from 15 May 2022 and the 81~ks observation from 12 June 2022. Our analysis involved the examination of Level--2 data, processed primary data collected by GPD through the instrumental pipeline, and we analyzed the data using \texttt{ixpeobssim} \citep{Baldini2022}, version 30.6.3. The extraction of source events was executed from a circular region with a radius of 150 arcsec centered on Cyg X--1, specifically in the 2--8~keV energy range, utilizing the \texttt{ixpeobssim/xpselect}. The observed region was notably luminous, rendering background effects negligible \citep{DiMarco2023}. 

From the two observation in May and June of 2022, we measured the polarization degrees and angles of $3.99\pm0.20$\%, $-21^\circ.3\pm1^\circ.4$ and $3.8\pm0.3$\%, $-25^{\circ}.4\pm2^\circ.3$, respectively by \texttt{ixpeobssim/xpbin PCUBE} algorithm \citep{Rankin2022,Kislat2015}. These values are consistent with the previous study (K22). The two observations are consistent in terms of the polarization information. K22 mentioned the May observation highlighted energy dependence. On the other hand, for the June observation, we measured the polarization degree was not significantly changed with energy as $3.7\pm0.3$\% at 2--4 keV, $4.0\pm0.5$\% at 4--6 keV and $3.8\pm1.1$\% at 6--8 keV. This disparity hints at variances in the origins of polarization, potentially tied to differences in accretion flows during each observation. Consequently, we approached the analysis of these observations, spaced a month apart, as distinct episodes and show the results of the June 2022 observation.

\section{Data Analysis and Results}
\begin{figure*}[tb]
	\begin{center}
	  \includegraphics[width=\linewidth]{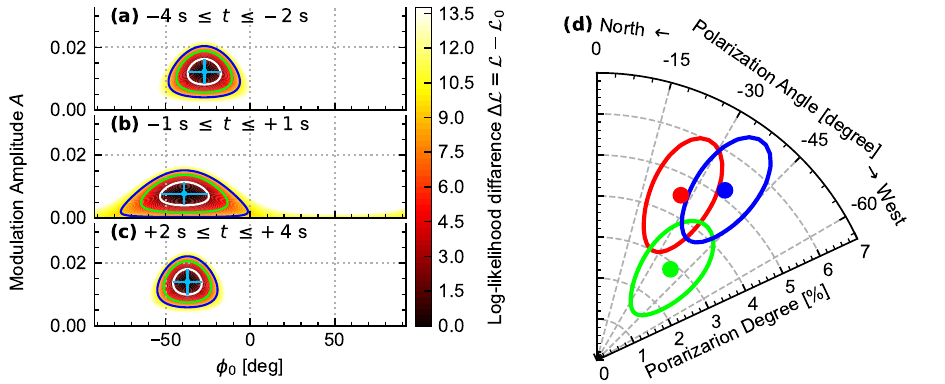} 
	\end{center}
	\caption{Time--segment resolved polarization as deduced from the modulation curve. (a)--(c), Log-likelihood contours for parameters $\phi_0$ and $A$ across distinct time segments: (a) ($-4~\mathrm{s} \leq t \leq -2~\mathrm{s}$), (b) ($-1~\mathrm{s} \leq t \leq +1~\mathrm{s}$), and (c) ($+2~\mathrm{s} \leq t \leq +4~\mathrm{s}$). The confidence intervals are represented as 68.3\% (white), 95.0\% (green), and 99.0\% (blue). Best--fit parameters are highlighted in cyan blue, with error bars signifying the $1\sigma$ (68.3\% confidence level). Contour shows the distribution of differences of log-likelihood $L$ from the minimum log-likelihood $L_0$, the log-likelihood value at the estimated $\phi_0$ and $A$. The minimum log-likelihood $L_0$ values for $-4~\mathrm{s} \leq t \leq -2~\mathrm{s}$, $-1~\mathrm{s} \leq t \leq +1~\mathrm{s}$, and $+2~\mathrm{s} \leq t \leq +4~\mathrm{s}$ intervals are 301.9, 305.2, and 290.0, respectively. (d) Polarization derived from assessed $\phi_0$ and $A$ values. Ellipses indicate the 68.3\% confidence regions. Results for time segments are color-coded as: $-4~\mathrm{s} \leq t \leq -2~\mathrm{s}$, (red), $-1~\mathrm{s} \leq t \leq +1~\mathrm{s}$ (green), and $+2~\mathrm{s} \leq t \leq +4~\mathrm{s}$ (blue).}
	\label{likelihood_and_polarplot}
\end{figure*}
  
\subsection{Analysis of aggregated X--ray intensity peaks}
We aggregated the peaks of the X--ray intensity obtained with IXPE light curve, referring to the previous shot analysis (N94; Y13). We created three light curves for each DU with a temporal bin size of 0.25~seconds, enabling the examination of variations below approximately 1~Hz. We first combined the light curves of two of the three DUs and used this combined light curve for the peak selection. Our criteria for peak selection involved assessing both the prominence and the inter-peak distances within the light curve. The prominence is the relative magnitude of the peaks and can efficiently eliminate small peaks. After testing various threshold settings, prominence showed to have a more critical influence on peak selection than inter-peak distance. We identified that a prominence of 60~counts/sec (corresponding to a 0.25-second bin) and an inter--peak distance of 8~s yielded the most statistically sound results and optimal polarization visibility (Figure \ref{lightcurve_and_shotprofile}(a)). At this step, this involves selectively identifying brighter peaks, including those misidentified due to Poisson noise. We therefore cross-referenced peak timings against the light curve from the uncombined DU, which was not utilized for peak selection process, and aggregated its photons. We identified 2546, 2825, and 2954 peaks for DU1, DU2, and DU3, respectively. Figure \ref{lightcurve_and_shotprofile}(b) displays the profile derived from these aggregated peaks and normalized using the average count rate observed from $-4$~s to $-2$~s and $+2$~s to $+4$~s around the peak.

\subsection{Time--segment resolved polarization}
We partitioned the stacked profile into seven time segments and determined the polarization within each time segment to comprehensively study the polarization changes across distinct peak segment. Time segments were set to $\pm 1$~s, centred on time points from $-3$~s to $+3$~s for 1~s each in the time relative to the peak time (time width of each segment was 2-second intervals). We identified events specific to each of these seven time segments from all three DUs and carried out polarization analysis.

For the determination of polarization, we employed three distinct methodologies. First, we utilized (1) \texttt{ixpeobssim/xpbin PCUBE} algorithm computational approach \citep{Rankin2022, Kislat2015}. Secondly, we conducted a (2) spectro--polarimetric fit using the \texttt{XSPEC} software (version v12.13.0c). Utilizing \texttt{XSPEC}, we simultaneously fitted the Stokes $I$, $Q$, and $U$ spectra from all three DUs for every time segment. We used the constant polarization model \texttt{POLCONST} for polarization model in \texttt{XSPEC}. Cyg X--1 was in the hard state in this observation, so the chosen model was \texttt{CONST*TBABS*POLCONST*(DISKBB+NTHCOMP)}. Specific parameters were fixed: column density in \texttt{TBABS} at $4\times10^{21} \mathrm{cm}^{-2}$, norm in \texttt{DISKBB} at 4000 and $kT_{\mathrm{e}}$ in \texttt{NTHCOMP} at 100~keV, and $kT_{\mathrm{bb}}$ was linked to \texttt{DISKBB}. As these fixed parameters are not determined by the IXPE band alone, we refer to K22, which reported the same hard state. The constant parameter is fixed to 1.0 at DU1, while those of the other two units are not fixed and determined as $0.9661\pm0.0012$ and $0.9197\pm0.0011$, respectively. Lastly, we performed (3) a modulation curve evaluation based on binned likelihood fitting (for more detailed in Appendix). A precedent \citet{HitomiCollaboration2018}, showcased the viability of employing this curve to distill X--ray polarization insights in a case of limited statistics. It is prudent to ascertain the consistency between statistical outcomes using \texttt{ixpeobssim/xpbin} and \texttt{XSPEC} and inferences drawn from the modulation curve. 

Figures \ref{likelihood_and_polarplot}(a)-(c) present the log--likelihood contours, derived from the likelihood estimation of the modulation curve's polarization amplitude $A$ and polarization angle $\phi_0$ for each three distinct intervals: $-4~\mathrm{s} \leq t \leq -2~\mathrm{s}$, $-1~\mathrm{s} \leq t \leq +1~\mathrm{s}$, and $+2~\mathrm{s} \leq t \leq +4~\mathrm{s}$. The log-likelihood within the time segment $-1~\mathrm{s} \leq t \leq +1~\mathrm{s}$ confirms polarization detection at a 95.5\% confidence level, whereas the other time segments record a robust 99.9\%. The time segment around the intensity peak, despite the higher number of photons before and after the peak segment, suggests that the anisotropy of photoelectrons changes not in any particular direction, but yielding the reduction of the modulation amplitude. As shown in Figure \ref{likelihood_and_polarplot}(d), polarizations derived from the modulation amplitude $A$ indicate that the polarization degree preceding and succeeding the maximum of the profile of the stacked peaks stands at $4.6\pm1.2$\% and $5.3\pm1.2$\% respectively. Interestingly, the polarization degree reduces to $3.4\pm1.1$\% at the peak (within a $\pm1$ second range). Concurrently, the polarization angles exhibit minor variability, hovering around $\sim30^\circ$. 

\begin{figure}[tb]
	\begin{center}
	  \includegraphics[width=\linewidth]{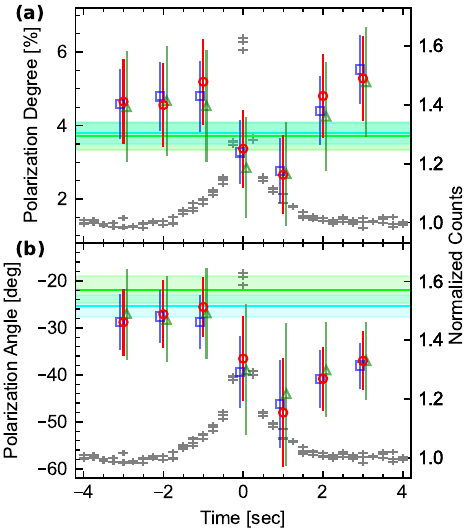} 
	\end{center}
	\caption{Intensity-correlated variability of polarization on a short--timescale. (a), Polarization degree and (b), Polarization angle, determined through three different analytical methods. Error bars indicate a 68.3\% confidence interval (1$\sigma$). Data are presented with a slight offset along the x--axis for visual clarity and are color-coded based on the analytical method used: \texttt{ixpeobssim/xpbin} (red circle), \texttt{XSPEC} (blue squared), and modulation curve (green triangle). For visual clarity, the normalized counts of intensity variability are presented on the two figures with the right axis. As a reference, the polarization averaged over the entire observation and outside the time segment of $-4 \leq t \leq +4$ are shown sin cyan and lime solid lines, where the patched regions indicate a 68.3\% confidence interval.}
	\label{PD_and_PA_with_shotprofile}
\end{figure}

{
\tabcolsep = 3.1pt
\begin{table*}[htbp]
	\caption{Polarization results (Polarization Degree (PD) and Polarization Angle (PA)) from three analysis methods and the best--fitting parameters across seven time segments. These results were derived using three distinct methods: \texttt{xpbin}, \texttt{XSPEC} and modulation curve.}
	\begin{tabular}{cccccccc}
		\hline\hline
		Time Segment~[s] & $-4 \leq t \leq -2$ & $-3 \leq t \leq -1$ & $-2 \leq t < 0$ & $-1 \leq t \leq 1$ & $0 \leq t \leq 2$ & $1 \leq t \leq 3$ & $2 \leq t \leq 4$ \\ 
		\hline
		PD$^{*}$~[\%] & $4.6\pm{1.2}$ & $4.6\pm{1.1}$ & $5.2\pm{1.2}$ & $3.4\pm{1.1}$ & $2.7\pm{1.1}$ & $4.8\pm{1.1}$ & $5.3\pm{1.2}$ \\

		PD$^{\dag}$~[\%]& $4.6\pm1.0$ & $4.8\pm1.0$ & $4.8\pm1.0$ & $3.3\pm0.9$ & $2.8\pm0.9$ & $4.4\pm1.0$ & $5.5\pm1.0$ \\

		$A/A_{100}$~[\%] & $4.5\pm1.5$ & $4.7\pm1.5$ & $4.5\pm1.5$ & $2.9\pm1.4$ & $2.7\pm1.4$ & $4.2\pm1.5$ & $5.2\pm1.5$ \\
		
		\hline
		PA$^{*}$~[deg] & $-29\pm{7}$ & $-27\pm{7}$ & $-26\pm{6}$ & $-37\pm{9}$ & $-48\pm{12}$ & $-40\pm{6}$ & $-37\pm{6}$ \\
		
		PA$^{\dag}$~[deg] & $-29\pm6$ & $-28\pm6$ & $-29\pm6$ & $-39\pm8$ & $-46\pm9$ & $-41\pm6$ & $-38\pm5$ \\

		$\phi_{0}$~[deg] & $-27\pm10$ & $-28\pm9$ & $-26\pm10$ & $-39\pm14$ & $-44\pm15$ & $-39\pm10$ & $-37\pm8$ \\

		\hline
		$T_{\mathrm{in}}$~[keV] & $0.352_{-0.020}^{+0.016}$ & $0.350_{-0.022}^{+0.017}$ & $0.353_{-0.024}^{+0.018}$ & $0.387_{-0.015}^{+0.012}$ & $0.382_{-0.015}^{+0.012}$ & $0.359_{-0.019}^{+0.015}$ & $0.365_{-0.017}^{+0.014}$ \\
		Gamma & $1.83_{-0.017}^{+0.016}$ & $1.836_{-0.017}^{+0.016}$ & $1.849_{-0.018}^{+0.017}$ & $1.823_{-0.018}^{+0.017}$ & $1.805_{-0.018}^{+0.017}$ & $1.812_{-0.017}^{+0.016}$ & $1.810_{-0.018}^{+0.017}$ \\
		$\chi^2$/d.o.f & $1401/1336$ & $1413/1336$ & $1440/1336$ & $1415/1336$ & $1392/1336$ & $1362/1336$ & $1369/1336$ \\ 
		\hline
		$F_{4-8}/F_{2-4}$ $^{\ddag}$ & $1.129\pm0.005$ & $1.127\pm0.005$ & $1.119\pm0.005$ & $1.119\pm0.005$ & $1.135\pm0.005$ & $1.141\pm0.005$ & $1.143\pm0.005$ \\ \hline\hline
	\end{tabular}
	\begin{tabnote}
		\footnotemark[$*$] Analysis method:\texttt{ixpeobssim/xpbin} \\ 
		\footnotemark[$\dag$] Analysis method:\texttt{XSPEC} \\
		\footnotemark[$\ddag$] un-absorbed flux ratio
	\end{tabnote}
	\label{Results_table}
\end{table*}
}

Figure \ref{PD_and_PA_with_shotprofile} displays the polarization degrees and the polarization angles for all time segment among the shot duration, obtained through the three independent methods (summarized in Table \ref{Results_table}). Our assessment of polarization approached via three distinct methodologies, yielded consistent outcomes. The polarization angle appears to be smaller than the time average. This is because the polarization angle is smaller in the extracted intensity variability, and the polarization angle outside the time segment of the intensity variability corresponds to the time average. When comparing segment $-2~\mathrm{s} \leq t < +0~\mathrm{s}$ and $+0~\mathrm{s} \leq t \leq +2~\mathrm{s}$ --- characterized by non--overlapping computed intervals ---, we discerned a shift in the polarization degree within a $1\sigma$ boundary. The polarization degree before the peak was $5.2\pm1.2$\% and after the peak decreased to $2.7\pm1.1$\%. The polarization angle also changed at the period immediately after the peak. While the statistical robustness of the findings is not yet optimal, the results suggest a potential temporal lag between polarization degree and intensity variability, where the polarization degree decreases on a timescale of seconds following the peak of the luminosity enhancement. 

\section{Discussion and Conclusion}
The observed short-timescale correlation with spectral changes, advances our comprehension of the accretion dynamics proximate to the black holes. The observed correlation among increase in X--ray intensity, softening of the spectrum and decrease in polarization is be able to interpret the observed luminosity increase as the inner edge of the standard accretion disk approaching the black hole (also reported in \citep{Bhargava2022}). The increase in unpolarized or low-polarized radiation due to the inner accretion disk edge being closer to the black hole may have softened the spectrum (the accretion disk component is at the lower energy side down to $\sim3$~keV) and reduced the total polarization (Table \ref{Results_table}). Also, following the context of stronger radiation originating from the accretion disk from the vicinity of the black hole, there is a difference in polarization angle between the direct radiation from the accretion disk and the scattered radiation in the corona. Consequently, the mixing of two or more polarization angle directions has the effect of reducing the polarization degree. This phenomenon explains the change in polarization angle and the reduction in the modulation of the modulation curve as shown in Figure \ref{likelihood_and_polarplot} and \ref{PD_and_PA_with_shotprofile}.

Intensity variability on the scale of a few seconds are related to mass accretion and can be compared with the structural changes associated with the transition from the low/hard to the high/soft state. \citet{JanaandChang2024} reported that they observed differences in the polarization of different hard and soft states of Cyg X--1. The authors suggested that the optical thickness of the corona in the soft state increased the scattering frequency of the seed photons and reduced the polarization. The decrease in polarization shown in this study can also be naturally explained by the density change due to the contraction of the corona near the black hole and the increase in the number of scattering of seed photons. Conventionally, regions with lower optical depth would exhibit less frequent scattering events, thus resulting in more pronounced polarization of Compton radiation \citep{Poutanen2023}, a scenario that diverges from our empirical findings. Also, there is no valid explanation for the change in polarization angle.

Additionally, other phenomena could be involved. For instance, it could be argued that the inflows and outflows in relatively short times affects the polarization\citep{Poutanen2023}, or else that the strong gravitational field of the black hole bends the light rays and rotates polarization. Moreover, the time lag between variability in polarization degree and angle and increase in X--ray intensity may represent a complex accretion process in the black hole binary. At the very least, the variability of polarization information suggests that a dynamic change in the accretion flow near the black hole. However, to verify this phenomenon, it is essential to observe polarization with good time resolution and sufficient statistics and a more detailed study of broader bandwidth polarized X--ray observations. 

For a deep understanding, it is desirable to investigate radiations across a wider polarization spectrum. By analyzing the rapid polarization variability in diverse astrophysical objects like black hole binaries, blazars, AGNs, and ultra--luminous sources, we can deepen our knowledge of the physics governing black holes. Furthermore, the technique of stacking analysis to detect polarization changes holds potential for broader application to other compact objects, such as accreting neutron stars or white dwarfs, exhibiting aperiodic intensity fluctuations.

\begin{ack}
This research used data products provided by the IXPE Team (MSFC, SSDC, INAF and INFN) and distributed with additional software tools by the High-Energy Astrophysics Science Archive Research Center (HEASARC), at NASA Goddard Space Flight Center (GSFC). This work was supported by JSPS KAKENHI Grant Number JP22H01272, JP20K20527, 22H01269, 21K1394, and 23K22540. 
\end{ack}

\appendix 
\section{Polarization Estimation Procedure through Modulation Curve}
The essence of polarization analysis lies in discerning the anisotropic distribution of photoelectron trajectories within the detector, called modulation curve. The GPDs of IXPE are designed to specialize in mapping this anisotropy, particularly through the azimuthal angle of photoelectron tracks during photoelectric absorption when the GPD detects polarized X--ray photons. Subsequently, polarization details are extracted by translating this anisotropy into the Stokes parameters, $Q$ and $U$, customized for each recorded event. 

To compute azimuthal angles of emitted photoelectrons for each event, designated as $k$ for $k$-th event, we applied the equation:
\begin{equation}
u_{k}\sim\sin(2\phi_k),~q_{k}\sim\cos(2\phi_k).
\end{equation}
Here, $u_{k}$, and $q_{k}$ represents the Stokes parameters for individual event and $\phi_{k}$ represents the azimuthal angle in sky coordinates. The symbol denoting approximately equal is used due to the concentration of photons from luminous sources, such as Cyg X--1, around the center of the detector, where systematic errors are minimized, in contrast to the peripheries of the detectors \citep{Rankin2022}. We constructed a histogram by using the azimuthal angle $\phi_{k}$, for every DU and subsequently aggregated the results from all units. To delineate the modulation curves and ascertain polarization properties, we employed a binned likelihood fit approach, according to methodologies presented in \citet{HitomiCollaboration2018}. 

Subsequently, we conducted two distinct simulation observations using the \texttt{ixpeobssim/xpobssim}: one reflecting an unpolarized observation and the other illustrating a 100\% polarized observation. These simulations were based on the Cyg X--1 radiation parameters specific to each DU. We set the spectral model for the simulation as \texttt{tbabs}$\times$\texttt{(diskbb+nthcomp)}. The parameters was determined by fitting the spectral data from each DU for the observation periods in June 2022, utilizing \texttt{XSPEC}. For our simulations, exposure times were designated as $10^4$ s for the unpolarized scenario and $10^5$ s for the completely polarized case. 

The expected counts for each $i$-th histogram bin, denoted as $n_{\mathrm{exp}} (\phi_i )$, is formulated as:
\begin{equation}
  n_{\mathrm{exp}}(\phi_i )=n_{\mathrm{sim}}(\phi_i )\{ 1+A\cos[2(\phi_i-\phi_0 )]\}.
\end{equation}
Here, $A$ represents the modulation curve's amplitude, $\phi_0$ is the polarization angle in sky coordinates. It is postulated that $n_{\mathrm{obs}}$ adheres to a Poisson distribution. The likelihood function is given by:
\begin{equation}
  L(\phi_0, A) = \prod_{i}\mathrm{Poisson}[n_{\mathrm{obs}}(\phi_i )/n_{\mathrm{exp}}(\phi_i )].
\end{equation}
And its logarithmic transformation can be described as:
\begin{equation}
	\mathcal{L}=-2\log L.
\end{equation}
To optimize the model, the maximum likelihood estimates for $\phi_{0}$ and $A$ are derived by minimizing the aforementioned logarithmic likelihood function. Denoting the minimal value of this function as $L_0$, the difference,  $\Delta \mathcal{L}=\mathcal{L}-\mathcal{L}_{0}$, converges to a $\chi^2$-distribution. For the two free parameters, $\phi_{0}$ and $A$, the $\Delta \mathcal{L}$ values of 2.30, 5.99, and 9.21 correspond to confidence levels of 68.3\%, 95.0\%, and 99.0\%, respectively. We showed the $\Delta \mathcal{L}$ distribution in Figure \ref{likelihood_and_polarplot}(a).

The polarization degree $\Pi$ was calculated using the following relation:
\begin{equation}
  \Pi = \frac{A}{A_{100}}
\end{equation}
Here, $A_{100}$ signifies the amplitude of the modulation curve derived from a 100\% polarized observation simulation. We evaluated the $A_{100}$ value as $0.270\pm0.001$ for the observations conducted in June 2022. The evaluated results are shown in Figure \ref{likelihood_and_polarplot}(b). We identified events specific from all three DUs to each of seven time segments and subsequently amalgamated their modulation curves to produce a singular comprehensive curve. An assessment of the modulation curves for each time-segment was carried out, with polarization values estimated via a binned likelihood fit.


\end{document}